\documentclass{tlp}
\usepackage{epsfig}
\usepackage[utf8]{inputenc}
\usepackage{multicol}
\usepackage{multirow}
\usepackage{wrapfig}
\usepackage{subfigure}
\usepackage{url}
\usepackage{verbatim}
\newcommand{\verbatimproperties}{\renewcommand{\baselinestretch}{0.85} \small}

\begin{document}

\label{firstpage}

\title[Instance Retrieval of Subgoals for Subsumptive Tabled Evaluation]
      {Efficient Instance Retrieval of Subgoals\\for Subsumptive Tabled Evaluation\\of Logic Programs}

\author[Flávio Cruz and Ricardo Rocha]
       {FLÁVIO CRUZ and RICARDO ROCHA\\
       CRACS \& INESC-Porto LA, Faculty of Sciences, University of Porto\\
       Rua do Campo Alegre, 1021/1055, 4169-007 Porto, Portugal\\
       \email{\{flavioc,ricroc\}@dcc.fc.up.pt}}

\maketitle


\begin{abstract}
  Tabled evaluation is an implementation technique that solves some
  problems of traditional Prolog systems in dealing with recursion and
  redundant computations. Most tabling engines determine if a tabled
  subgoal will produce or consume answers by using variant checks. A
  more refined method, named call subsumption, considers that a
  subgoal $A$ will consume from a subgoal $B$ if $A$ is subsumed by
  (an instance of) $B$, thus allowing greater answer reuse. We
  recently developed an extension, called \emph{Retroactive Call
    Subsumption}, that improves upon call subsumption by supporting
  bidirectional sharing of answers between subsumed/subsuming
  subgoals. In this paper, we present both an algorithm and an
  extension to the table space data structures to efficiently
  implement instance retrieval of subgoals for subsumptive tabled
  evaluation of logic programs. Experiments results using the YapTab
  tabling system show that our implementation performs quite well on
  some complex benchmarks and is robust enough to handle a large
  number of subgoals without performance degradation.
\end{abstract}



\section{Introduction}

Tabled resolution methods solve some of the shortcomings of Prolog
because they can reduce the search space, avoid looping and have
better termination properties than SLD resolution
methods~\cite{Chen-96}. Tabling works by memorizing generated answers
and then by reusing them on \emph{similar calls} that appear during
the resolution process. In a nutshell, first calls to tabled subgoals
are considered \emph{generators} and are evaluated as usual, using SLD
resolution, but their answers are stored in a global data space,
called the \emph{table space}. Similar calls are called
\emph{consumers} and are resolved by consuming the answers already
stored for the corresponding generator, instead of re-evaluating them
against the program clauses. There are two main approaches to
determine if two subgoals $A$ and $B$ are similar:

\begin{itemize}
\item \emph{Variant-based tabling}: $A$ and $B$ are variants if they
  can be made identical by variable renaming. For example, $p(X,1,Y)$
  and $p(Y,1,Z)$ are \emph{variants} because both can be transformed
  into $p(VAR_0,1,VAR_1)$;
\item \emph{Subsumption-based tabling}: $A$ is considered similar to
  $B$ if $A$ is \emph{subsumed} by $B$ (or $B$ \emph{subsumes} $A$),
  i.e., if $A$ is more specific than $B$ (or an instance of).  For
  example, subgoal $p(X,1,2)$ is subsumed by subgoal $p(Y,1,Z)$
  because there is a substitution $\{Y=X,Z=2\}$ that makes $p(X,1,2)$
  an instance of $p(Y,1,Z)$. Tabling by call subsumption is based on
  the principle that if $A$ is subsumed by $B$ and $S_A$ and $S_B$ are
  the respective answer sets, then $S_A \subseteq S_B$. Note that this
  assumption may not hold when programs use extra-logical features of
  Prolog such as the \texttt{var/1} and \texttt{nonvar/1} built-in
  predicates.
\end{itemize}

For some types of programs, subsumption-based tabling yields superior
time performance, as it allows greater reuse of answers, and better
space usage, since the answer sets for the subsumed subgoals are not
stored. However, the mechanisms to efficiently support
subsumption-based tabling are harder to implement, which makes
subsumption-based tabling not as popular as variant-based tabling. XSB
Prolog~\cite{Rao-97} was the first Prolog system to implement
subsumption-based tabling, first by using a data structure called
\emph{Dynamic Threaded Sequential Automata (DTSA)}~\cite{Rao-96} and
later by using a data structure called \emph{Time-Stamped Trie
  (TST)}~\cite{Johnson-99,Johnson-00}, that showed better space
efficiency than DTSA. Despite the advantages of using
subsumption-based tabling, the degree of answer reuse might depend on
the call order of subgoals. For example, in XSB, if a more general
subgoal is called before specific subgoals, answer reuse will happen,
but if more specific subgoals are called before a more general
subgoal, no reuse will occur.

In order to solve this problem, we implemented an extension to the
original TST design, called \emph{Retroactive Call Subsumption
  (RCS)}~\cite{Cruz-10}, that supports subsumption-based tabling by
allowing full sharing of answers between subsumptive subgoals,
independently of the order they are called. RCS works by selectively
pruning the evaluation of subsumed subgoals when a more general
subgoal appears later on. We designed a new table space organization,
called \emph{Single Time-Stamped Trie (STST)}~\cite{Cruz-10} where
answers are represented only once. We also designed a new algorithm to
efficiently retrieve the set of currently evaluating subgoals that are
subsumed by (instances of) a more general subgoal. In this paper, we
will focus our discussion on the support for this algorithm on a
concrete implementation, the YapTab system~\cite{Rocha-05a,Rocha-00a},
but our proposals can be generalized and applied to other tabling
systems based on similar data structures.

The remainder of the paper is organized as follows. First, we briefly
introduce the main background concepts about the table space in
YapTab. Next, we describe the modifications made to the table space
data structures and we discuss in detail our new algorithm to
efficiently retrieve subsumed subgoals. Finally, we present some
experimental results and then we finish by outlining some conclusions.


\section{Table Space}

\begin{figure}[ht]
\centering
\includegraphics[scale=0.50]{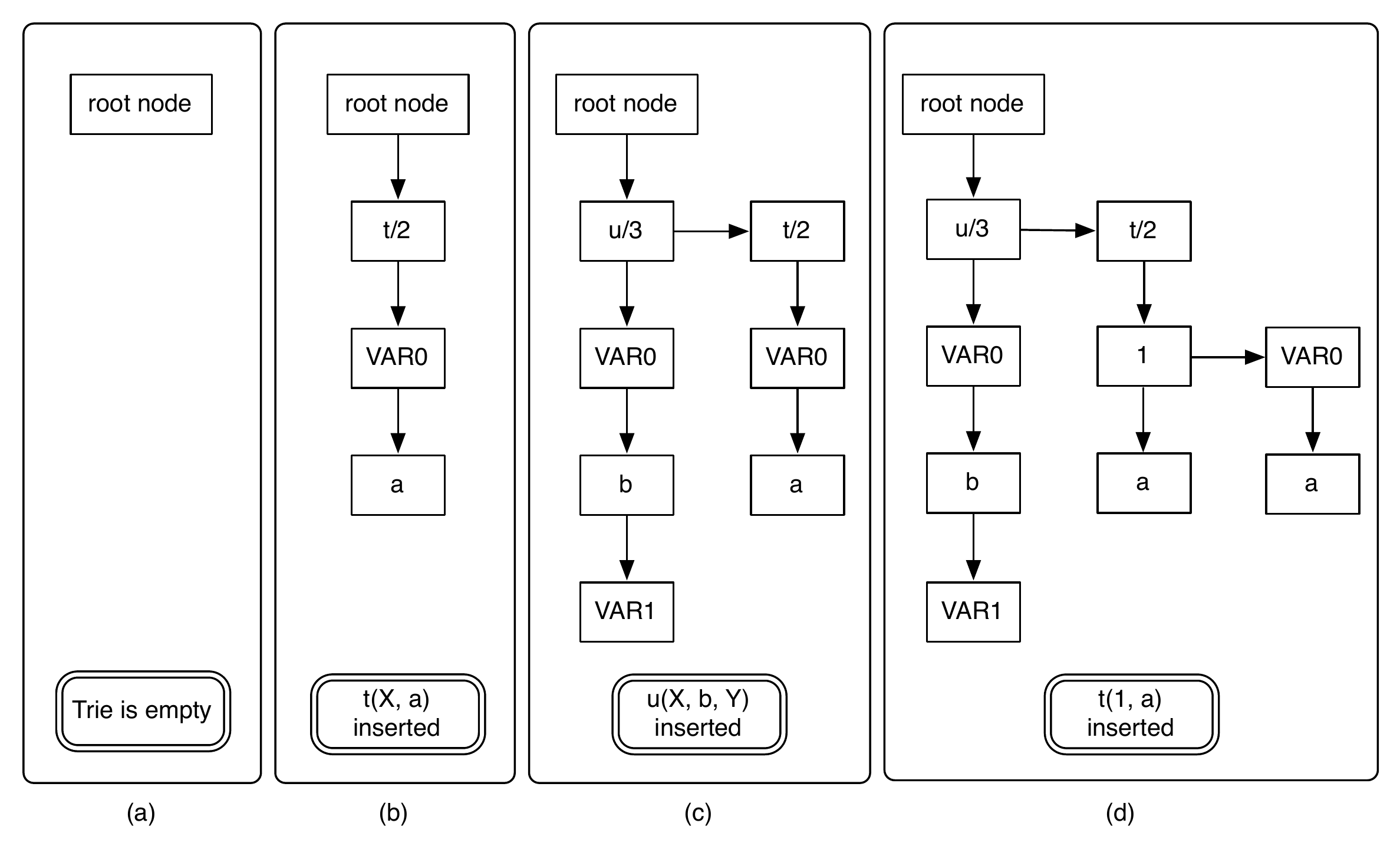}
\caption{Using tries to represent terms}
\label{fig:tries_use}
\end{figure}

Whenever a tabled subgoal is first called, a new entry is allocated in
the table space. Table entries are used to keep track of subgoal calls
and to store their answers. Arguably, the most successful data
structure for tabling is \emph{tries}~\cite{RamakrishnanIV-99}. Tries
are trees in which common prefixes are represented only once. Tries
provide complete discrimination for terms and permit look-up and
insertion to be done in a single pass. Figure~\ref{fig:tries_use}
shows an example of a trie. First, in (a) the trie is only represented
by a \textit{root node}. Next, in (b) the term $t(X,a)$ is inserted
and three nodes are created to represent each part of the term. In (c)
a new term, $u(X,b,Y)$ is inserted.  This new term differs from the
first one and a new distinct branch is created. Finally, in (d), the
term $t(1,a)$ is inserted and only one new node needs to be created as
this term shares the two prefix nodes with term $t(X,a)$. Note that
$t(1,a)$ is subsumed by $t(X,a)$ since there is a substitution, namely
$\{X=1\}$, that makes $t(1,a)$ an instance of $t(X,a)$.

In YapTab, we use the \emph{trie node} data structure as the building
block of tries. Each node contains the following fields: $symbol$,
representing the stored part of a term; $child$, a pointer to the
child node; $parent$, a pointer to the parent node; $sibling$, which
points to the sibling node; and $status$, a bit field with information
about the trie node.  When the chain of sibling nodes exceeds a
certain threshold length, a hashing scheme is dynamically employed to
provide direct access to nodes, optimizing the search of terms.

In the STST organization, each tabled predicate has a \emph{table
  entry} that points to a \emph{subgoal trie} and an \emph{answer
  trie}. In the subgoal trie, each distinct trie path represents a
tabled subgoal call and each leaf trie node points to a \emph{subgoal
  frame}, a data structure containing information about the subgoal
call, notably an \emph{answer return list} containing pointers to leaf
nodes of the answer trie. In order to support call subsumption
mechanisms, the answer trie is also a \emph{time-stamped
  trie}~\cite{Johnson-99}, where the trie nodes are extended with a
$timestamp$ field.


\section{Retrieval of Subsumed Subgoals}

In this section, we describe the modifications made to the subgoal
trie data structure and we discuss in detail the new algorithm
developed to efficiently retrieve the set of currently evaluating
instances of a subgoal. Note that in a RCS setting, this new algorithm
is executed only when the currently executing subgoal is a generator
(and not a consumer of answers) and a call represents a generator only
if no variant or subsuming subgoals are found during an initial search
on the subgoal trie. This initial search is done in a single pass
through the subgoal trie using the method initially proposed
by~\cite{Rao-96} for non-retroactive subsumption-based tabling.


\subsection{Subgoal Trie Data Structure}

\begin{wrapfigure}{!r}{7cm}
\centering
\includegraphics[scale=0.45]{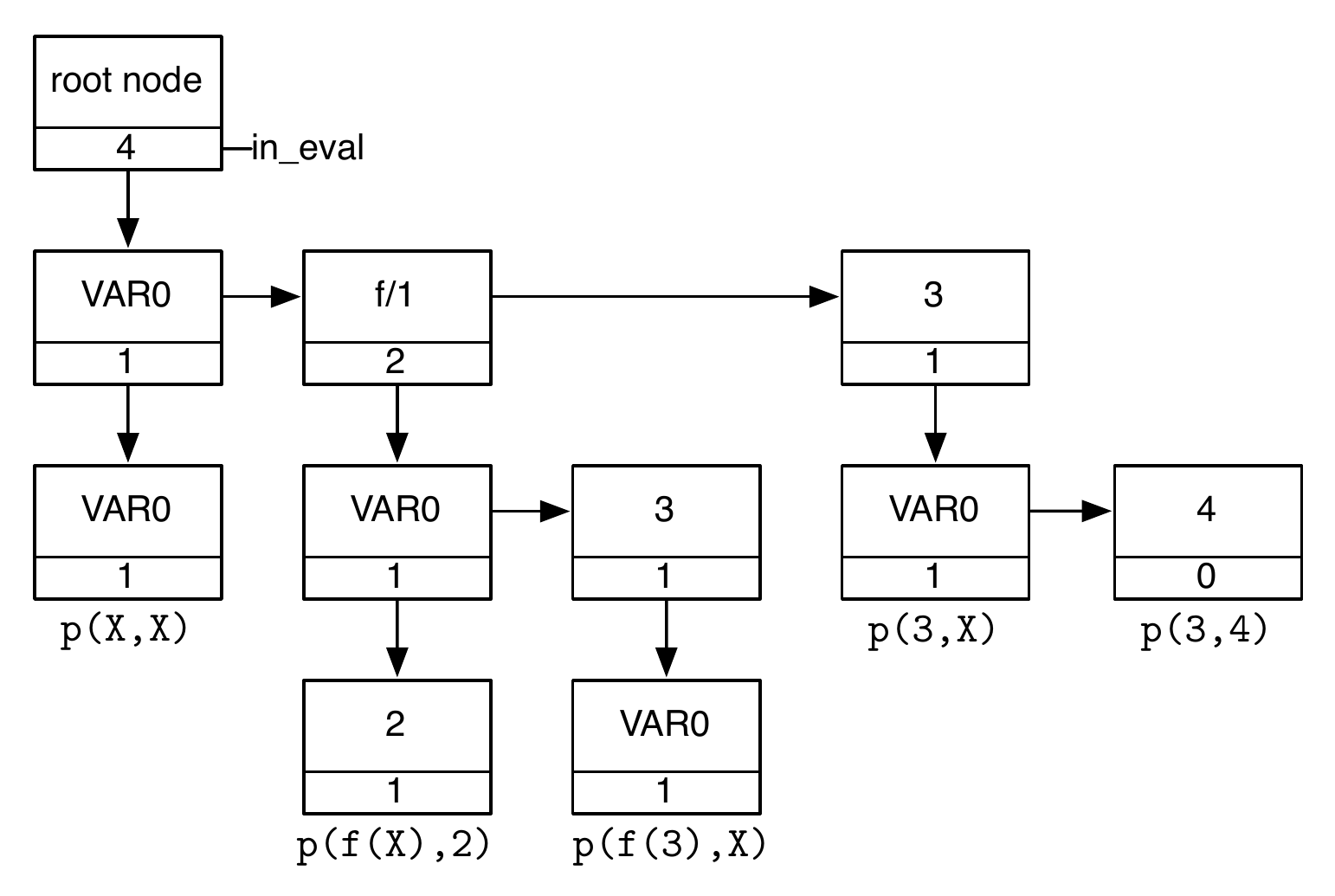}
\caption{The $in\_eval$ field in a subgoal trie representing a $p/2$
  tabled predicate}
\label{fig:in_eval_trie}
\end{wrapfigure}

Each subgoal trie node was extended with a new field, named
$in\_eval$, which stores the number of subgoals, represented below the
node, that are in evaluation. During the search for subsumed subgoals,
this field is used to ignore the subgoal trie branches without
evaluating subgoals, i.e., the ones with $in\_eval = 0$.

When a subgoal starts being evaluated, all subgoal trie nodes in its
subgoal trie path get the $in\_eval$ field incremented. When a subgoal
completes its evaluation, the corresponding $in\_eval$ fields are
decremented. Hence, for each subgoal leaf trie node, the $in\_eval$
field can be equal to either: 1, when the corresponding subgoal is in
evaluation; or 0, when the subgoal is completed. For the root subgoal
trie node, we know that it will always contain the total number of
subgoals being currently evaluated. For an example, the subgoal trie
in Fig.~\ref{fig:in_eval_trie} represents four evaluating subgoals and
one completed subgoal for the tabled predicate $p/2$.

When a chain of sibling nodes is organized in a linked list, it is
easy to select the trie branches with evaluating subgoals by looking
for the nodes with $in\_eval > 0$. But, when the sibling nodes are
organized in a hash table, it can become very slow to inspect each
node as the number of siblings increase. In order to solve this
problem, we designed a new data structure, called \textit{evaluation
  index}, in a similar manner to the \emph{timestamp
  index}~\cite{Johnson-99} of the TST design.

An evaluation index is a double linked list that is built for each
hash table and is used to chain the subgoal trie nodes where the
$in\_eval$ field is greater than 0. Note that this linked list is not
ordered by the $in\_eval$ value. Each evaluation index node contains
the following fields: $prev$, a pointer to the previous evaluation
index node, if any; $next$, a pointer to the next evaluation index
node, if any; $node$, a pointer to the subgoal trie node the index
node represents; and $in\_eval$, the number of evaluating subgoals
under the corresponding subgoal trie node. We also extended the hash
table with a field named $index$ to point to the evaluation index.

\begin{figure}[ht]
\centering
\includegraphics[scale=0.50]{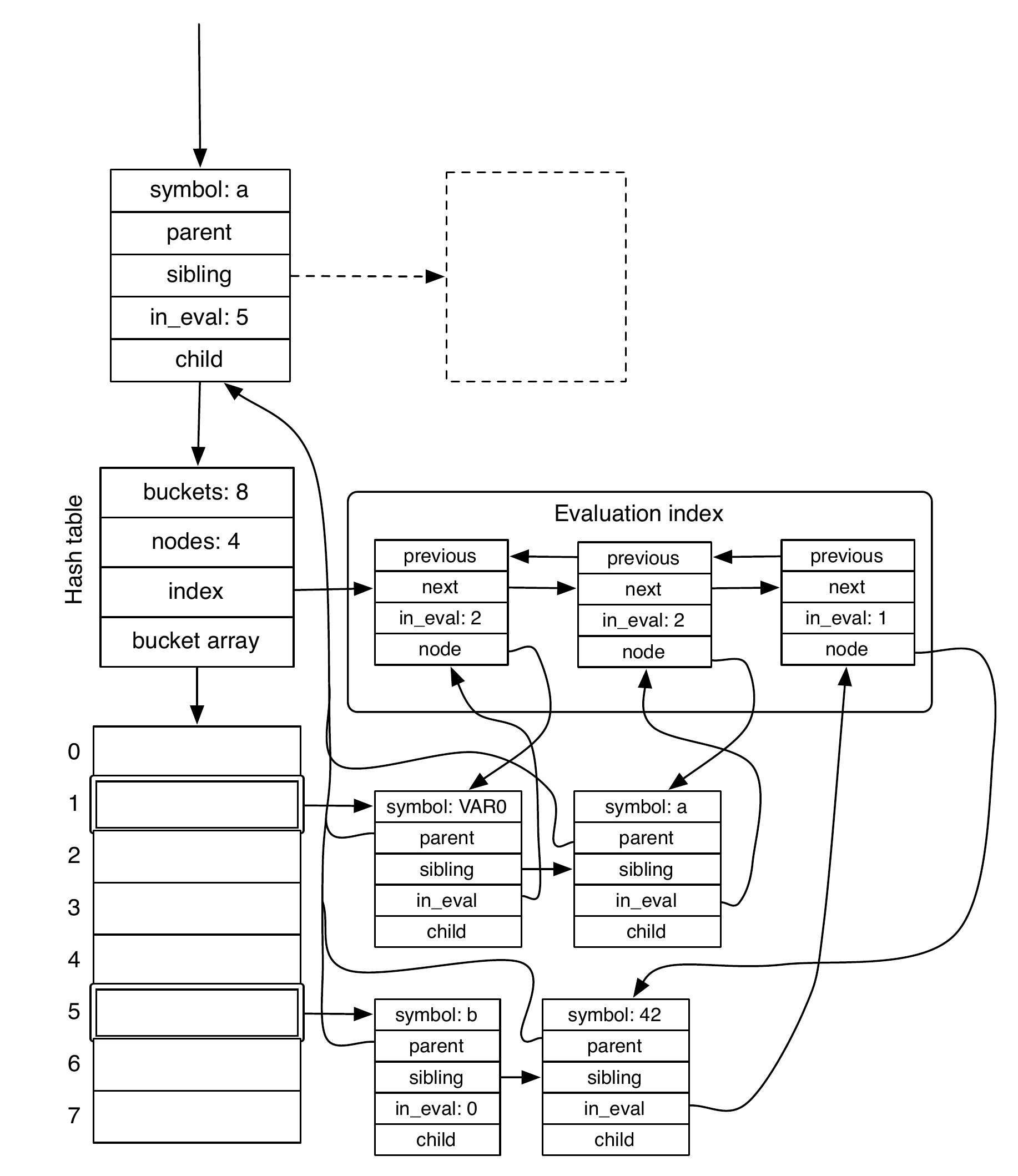}
\caption{A hash table with an evaluation index}
\label{fig:hash_table_evaluation_index}
\end{figure}

An indexed subgoal trie node uses the $in\_eval$ field to point to the
index node, while a trie node with $in\_eval = 0$ is not indexed. To
compute the $in\_eval$ value of a trie node, we first need to use the
$status$ field to determine if the node is inside a hash table or not,
and then use the $in\_eval$ field
accordingly. Figure~\ref{fig:hash_table_evaluation_index} shows a hash
table and the corresponding evaluation index.

The evaluation index makes the operation of pruning trie branches more
efficient by providing direct access to trie nodes with evaluating
subgoals, instead of visiting a potentially large number of trie nodes
in the hash table. While advantageous, the operation of incrementing
or decrementing the $in\_eval$ fields in a trie path is more costly,
because these indexes must be maintained.

Figure~\ref{fig:increment_in_eval} presents the pseudo-code for the
$increment\_in\_eval()$ procedure. This procedure iterates over the
subgoal trie path and increments the $in\_eval$ field from the leaf to
the root node. When we find hashed trie nodes, we must check if the
node is currently being indexed. If this is the case, we simply
increment the $in\_eval$ field of the index node, otherwise we create
a new index node on the evaluation index pointing to the current
subgoal trie node and the $in\_eval$ field of the subgoal trie node is
made to point to the index node.

\begin{figure}[!ht]
\verbatimproperties
\begin{verbatim}
increment_in_eval(leaf_node, root_node) {
  current_node = leaf_node
  while (current_node != root_node)
    if (is_hashed_node(current_node))
      if (in_eval(current_node) == 0)                         // not indexed
        hash_table = child(parent(current_node))
        index_node = add_index_node(hash_table, current_node)
        in_eval(current_node) = index_node
      else                                                        // indexed
        index_node = in_eval(current_node)
        in_eval(index_node) = in_eval(index_node) + 1
    else                                                // simple chain list
      in_eval(current_node) = in_eval(current_node) + 1
    current_node = parent(current_node)
  in_eval(root_node) = in_eval(root_node) + 1
}
\end{verbatim}
\caption{Pseudo-code for procedure $increment\_in\_eval()$}
\label{fig:increment_in_eval}
\end{figure}

The procedure in Fig.~\ref{fig:decrement_in_eval},
$decrement\_in\_eval()$, does the inverse job of procedure
$increment\_in\_eval()$. When decrementing an indexed subgoal trie
node, if the $in\_eval$ field reaches 0, the trie node no longer needs
to be indexed, and hence we must remove the index node from the
evaluation index. In the other cases, we simply decrement the
respective $in\_eval$ field.

\begin{figure}[!ht]
\verbatimproperties
\begin{verbatim}
decrement_in_eval(leaf_node, root_node) {
  current_node = leaf_node
  while (current_node != root_node)
    if (is_hashed_node(current_node))
      index_node = in_eval(current_node)
      if (in_eval(index_node) == 1)                     // remove from index
        hash_table = child(parent(current_node))
        remove_index_node(hash_table, index_node)
        in_eval(current_node) = 0
      else                                                   // keep indexed
        in_eval(index_node) = in_eval(index_node) - 1
    else                                                // simple chain list
      in_eval(current_node) = in_eval(current_node) - 1
    current_node = parent(current_node)
  in_eval(root_node) = in_eval(root_node) - 1
}
\end{verbatim}
\caption{Pseudo-code for procedure $decrement\_in\_eval()$}
\label{fig:decrement_in_eval}
\end{figure}


\subsection{Matching Algorithm}

The algorithm that finds the currently running subgoals that are
subsumed by a more general subgoal $S$ works by matching the subgoal
arguments $SA$ of $S$ against the trie symbols in the subgoal trie
$T$. By using the $in\_eval$ field as described previously, we can
ignore irrelevant branches as we descend the trie. When reaching a
leaf node, we append the corresponding subgoal frame in a result list
that is returned once the process finishes. If the matching process
fails at some point or if a leaf node was reached, the algorithm
backtracks to try alternative branches, in order to fully explore the
subgoal trie $T$.

When traversing $T$, trie variables cannot be successfully matched
against ground terms of $SA$. Ground terms of $SA$ can only be matched
with ground terms of $T$. For example, if matching the trie subgoal
$p(VAR_0,VAR_1)$ with the subgoal $p(2,X)$, we cannot match the
constant $2$ against the trie variable $VAR_0$, because $p(2,X)$ does
not subsume $p(X,Y)$.

When a variable of $SA$ is matched against a ground term of $T$,
subsequent occurrences of the same variable must also match the same
term. As an example, consider the trie subgoal $p(2,4)$ and the
subgoal $p(X,X)$. The variable $X$ is first matched against 2, but the
second matching, against 4, must fail because $X$ is already bound to
2.

Now consider the trie subgoal $p(VAR_0,VAR_1)$ and the subgoal
$p(X,X)$. Variable $X$ is first matched against $VAR_0$, but then we
have a second match against a different trie variable, $VAR_1$. Again,
the process must fail because $p(X,X)$ does not subsume $p(X,Y)$. This
last example evokes a new rule for variable matching. When a variable
of $SA$ is matched against a trie variable, subsequent occurrences of
the same variable must always match the same trie variable. This is
necessary, because the subgoals found must be \emph{instances} of
$S$. To implement this algorithm, we use the following data
structures:

\begin{itemize}
\item \textit{WAM data structures}: heap, trail, and associated
  registers. The heap is used to build structured terms, in which the
  subgoal arguments are bound. Whenever a new
  variable is bound, we trail it using the WAM trail;
\item \textit{term stack}: stores the remaining terms to be matched
  against the subgoal trie symbols;
\item \textit{term log stack}: stores already matched terms from the
  term stack and is used to restore the state of the term stack when
  backtracking;
\item \textit{variable enumerator vector}: used to mark the term
  variables that were matched against trie variables;
\item \textit{choice point stack}: stores choice point frames, where
  each frame contains information needed to restore the computation in
  order to search for alternative branches.
\end{itemize}

Figure~\ref{fig:collect_subsumed_subgoals} shows the pseudo-code for
the procedure that traverses a subgoal trie and collects the set of
subsumed subgoals of a given subgoal call. This procedure can be
summarized in the following steps:

\begin{enumerate}
\item setup WAM machinery and define the current trie node as the trie
  root node;
\item push the subgoal arguments into the term stack from right to
  left, so that the leftmost argument is on the top;
\item fetch a term $T$ from the term stack;
\item search for a child trie node $N$ of the current node where the
  $in\_eval$ field is not 0.
\item search for the next child node with a valid $in\_eval$ field to
  be pushed on the choice point stack, if any;
\item match $T$ against the trie symbol of $N$;
\item proceed into the child of $N$ or, if steps 4 or 6 fail,
  backtrack by popping a frame from the choice point stack. The frame
  is used to restore the term stack from the term log stack and to
  set the current trie node to the alternative node;
\item once a leaf is reached, add the corresponding subgoal frame to
  the resulting subgoal frame list. If there are choice points
  available, backtrack to try them;
\item if no more choice point frames exist, return the found subsumed
  subgoals.
\end{enumerate}

\begin{figure}[ht]
\verbatimproperties
\begin{verbatim}
collect_subsumed_subgoals(subgoal_trie, subgoal_call) {
  save_wam_registers()
  subgoals = NULL
  parent = subgoal_trie
  node = child(parent)
  push_arguments(term_stack, subgoal_call)
  while (true)
    term = deref(pop(term_stack))    
    if (is_atom(term) or is_integer(term))
      try_node = try_constant_term(term, node)
    else if (is_functor(term) or is_list(term))
      try_node = try_structured_term(term, node)
    else                                          // term must be a variable
      try_node = try_variable_term(term, node)
    if (try_node != NULL)
      push(term_log_stack, term)
      parent = try_node
      node = child(parent)
      if (empty(term_stack))                   // new subsumed subgoal found
        add_subgoal(subgoals, subgoal_frame(parent))
      else
        continue
    if (empty(choice_point_stack))
      unwind_wam_trail()
      restore_wam_registers()
      return subgoals
    else
      frame = pop_choice_point_frame(choice_point_stack)
      restore_computation(frame)
      node = get_alt_node(frame)
      parent = parent(node)
}
\end{verbatim}
\caption{Pseudo-code for procedure $collect\_subsumed\_subgoals()$}
\label{fig:collect_subsumed_subgoals}
\end{figure}


\subsection{Choice Point Stack}

To store alternative branches for exploration, we use a choice point
stack. Each choice point frame stores the following fields:
$alt\_node$, the alternative node to explore; $term\_stack\_top$, the
top of the term stack; $term\_log\_stack\_top$, the top of the term
log stack; $trail\_top$, the current trail position; and $saved\_HB$,
the $HB$ register. The $HB$ register is used to detect
\emph{conditional bindings} in the same manner as for the $HB$
register in WAM choice points, that is, we use it to know if a term
variable needs to be trailed. When a choice point frame is popped from
the stack, the state of the computation is restored by executing the
following actions:

\begin{itemize}
\item all terms stored in the term log stack are pushed back to the
  term stack in the inverse order (topmost terms are pushed first);
\item the trail is unwound to reset the variables that were bound
  after choice point creation;
\item register $H$ and $HB$ are reset to their previous values. $H$ is
  set to $HB$ and $HB$ to $saved\_HB$;
\item the current node and parent node are reset.
\end{itemize}

Since constant and structured terms can have at most one matching
alternative in a trie level, choice point frames are only pushed when
the current term is a variable. Remember that if a node satisfies the
$in\_eval$ prerequisite, variable terms can match all types of trie
symbols, including trie variables.


\subsection{Matching Constant and Structured Terms}

If the next term from the term stack is a constant or a structured
term we must match it against a similar ground term only. Both
constant and structured terms work pretty much the same way, except
that for a list or a functor term we push the term arguments into the
term stack before descending into the next trie level. The arguments
are pushed into the stack in order to be matched against the next trie
symbols. Figure~\ref{fig:try_structured_term} presents the pseudo-code
for the $try\_structured\_term()$ procedure.

\begin{figure}[ht]
\verbatimproperties
\begin{verbatim}
try_structured_term(term, current_node)
  if (is_hash_table(current_node))              // step 1: check hash bucket
    hash_table = current_node
    current_node = search_bucket_array(hash_table, term)
    if (current_node == NULL)
      return NULL
  while (current_node != NULL)    // step 2: traverse chain of sibling nodes
    if (symbol(current_node) == principal_functor(term))
      if (in_eval(current_node) > 0)
        push_arguments(term_stack, term)
        return current_node
      else                                      // no running subgoals below
        return NULL
    current_node = sibling(current_node)
  return NULL
}
\end{verbatim}
\caption{Pseudo-code for procedure $try\_structured\_term()$}
\label{fig:try_structured_term}
\end{figure}

This procedure is divided into two steps. In step 1, we check if the
current node is a hash table. If this is the case, we hash the term to
get the hash bucket that might contain the matching trie node. If the
bucket is empty we simply return $NULL$. Otherwise, we move into step
2. In step 2 we traverse a chain of sibling nodes (a simple chain or a
bucket chain) looking for a node with a matching symbol and with a
valid $in\_eval$ value.


\subsection{Matching Variable Terms}

A variable term can potentially be matched against any trie symbol. It
is only when the variable is matched against a trie variable that the
process may fail. Figure~\ref{fig:try_variable_term} shows the
pseudo-code for the $try\_variable\_term()$ procedure. It is defined
by three main cases, depending on the type of the current node,
namely:

\begin{enumerate}
\item the node is a hash table. For faster access of valid trie
  nodes, we use the evaluation index, which gives us all the valid
  trie nodes in a linked list. We set the next alternative node to be
  pushed on the choice point stack by using the function
  $next\_valid\_index\_node()$ that uses the $next$ pointer of the
  first index node to locate the alternative trie node.
\item the node is a hashed node, thus is on the evaluation index of
  the corresponding hash table. In this case, we also use the
  $next\_valid\_index\_node()$ function to identify the next
  alternative trie node.
\item the node is part of a simple linked list. Here we must use the
  function $next\_valid\_node()$ to find the next valid trie node
  ($in\_eval > 0$). The alternative trie node is also set using this
  function on the sibling node.
\end{enumerate}

\begin{figure}[ht]
\verbatimproperties
\begin{verbatim}
try_variable_term(variable, current_node) {
  if (is_hash_table(current_node))                     // case 1: hash table
    hash_table = current_node
    index_node = index(hash_table)    
    if (index_node == NULL)                     // no running subgoals below
      return NULL    
    current_node = node(index_node)
    alt_node = next_valid_index_node(current_node)
  else if (is_hashed_node(current_node))             // case 2: indexed node
    alt_node = next_valid_index_node(current_node)
  else                                         // case 3: simple linked list
    current_node = next_valid_node(current_node)
    if (current_node == NULL)
      return NULL
    alt_node = next_valid_node(current_node)
  push_choice_point_frame(choice_point_stack, {alt_node, ..., HB})
  if (try_variable_matching(variable, symbol(current_node)))
    return current_node
  else
    return NULL
}
\end{verbatim}
\caption{Pseudo-code for procedure $try\_variable\_term()$}
\label{fig:try_variable_term}
\end{figure}

After the current valid node and alternative node are set, we push the
alternative into the choice point stack and call the
$try\_variable\_matching()$ procedure
(Fig.~\ref{fig:try_variable_matching}) to match the term variable with
the trie node symbol.

\begin{figure}[ht]
\verbatimproperties
\begin{verbatim}
try_variable_matching(variable, symbol) {
  if (is_variable(symbol))
    if (is_in_variable_enumerator_vector(variable))
      return (variable_enumerator_index(variable) == var_index(symbol))
    else                                               // free term variable
      var_index = var_index(symbol)
      mark_variable_enumerator_vector(variable, var_index)
      return TRUE
  else                                                 // ground trie symbol
    if (is_in_variable_enumerator_vector(variable))
      return FALSE
    if (is_constant(symbol))
      bind_and_conditionally_trail(variable, symbol)
    else if (is_functor(symbol) or is_list(symbol))
      term = create_heap_structure(symbol)
      bind_and_conditionally_trail(variable, term)
      push_arguments(term_stack, term)
    return TRUE
}
\end{verbatim}
\caption{Pseudo-code for procedure $try\_variable\_matching()$}
\label{fig:try_variable_matching}
\end{figure}

Matching a term variable with a trie symbol depends on the type of the
trie symbol. If the trie symbol is a trie variable, we have two cases.
If the term variable is free (i.e., this is its first occurrence), we
simply make it to point to the position on the variable enumerator
vector that corresponds to the trie variable index and we trail the
term variable using the WAM trail. Otherwise, the term variable is
already matched against a trie variable (on the variable enumerator
vector), thus we get both indexes (term and trie variable indexes) and
the matching succeeds if they correspond to the same index (same
variable).

If the trie symbol is a ground term, we first test if the term
variable is on the variable enumerator vector and, in such case, we
fail since term variables matched against trie variables must only be
matched against the same trie variable. For constant trie symbols, we
simply bind the term variable to the trie symbol. For structured terms
(lists and functors), we create the structured term on the heap, bind
the term variable to the heap address, and push the new term arguments
into the term stack to be matched against the next trie symbols.


\subsection{Running Example}

\begin{figure}[ht]
\centering
\includegraphics[scale=0.45]{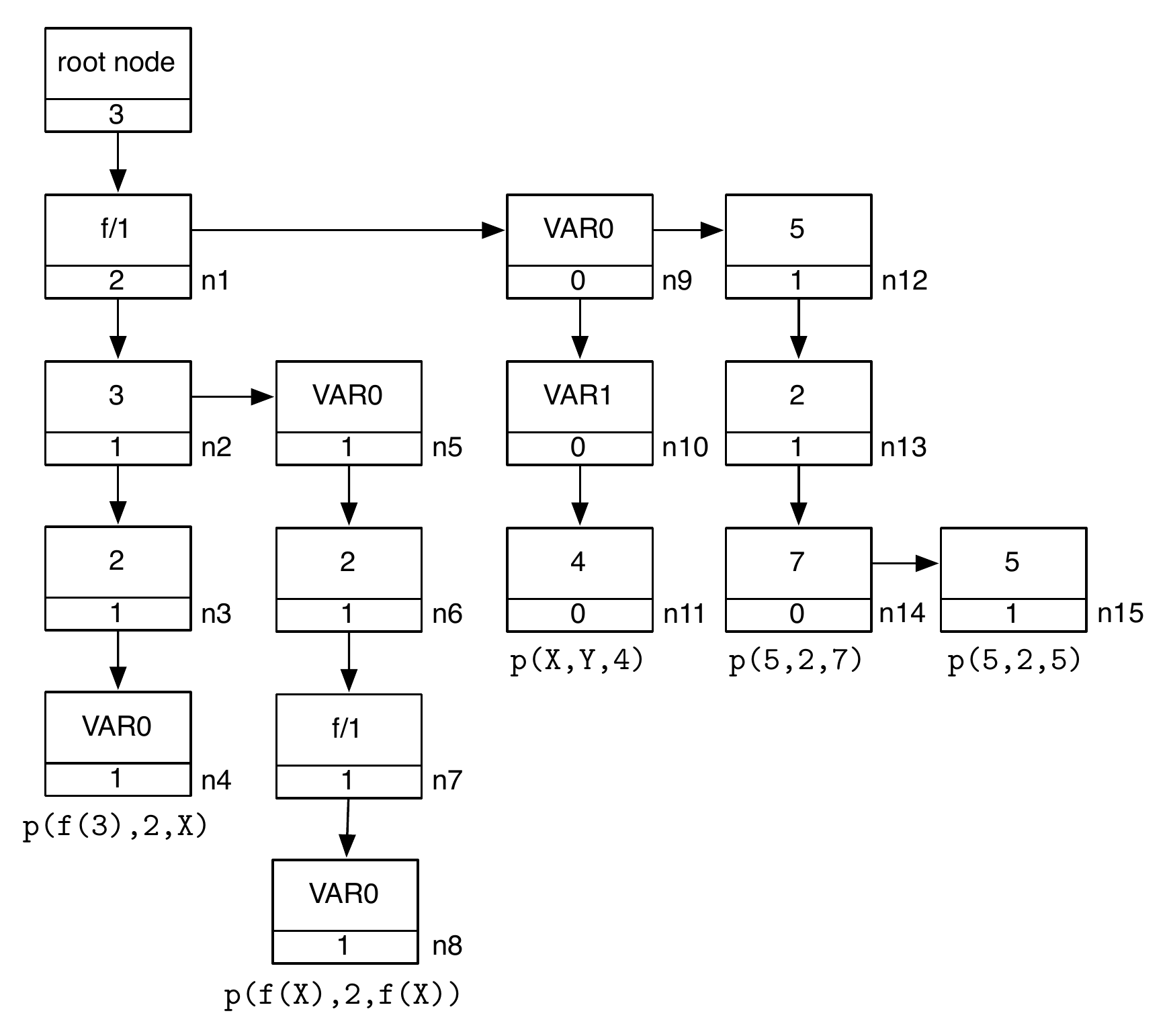}
\caption{A subgoal trie for a $p/3$ tabled predicate with three
  evaluating subgoals}
\label{fig:example_trie}
\end{figure}

Consider the subgoal trie in Fig.~\ref{fig:example_trie} representing
three evaluating subgoals and two completed subgoals for a tabled
predicate $p/3$. Now assume that we want to retrieve the subgoals that
are subsumed by the subgoal call $p(X,2,X)$.

\begin{figure}[ht]
\centering
\subfigure[Before descending into node $n2$]{\includegraphics[scale=0.42]{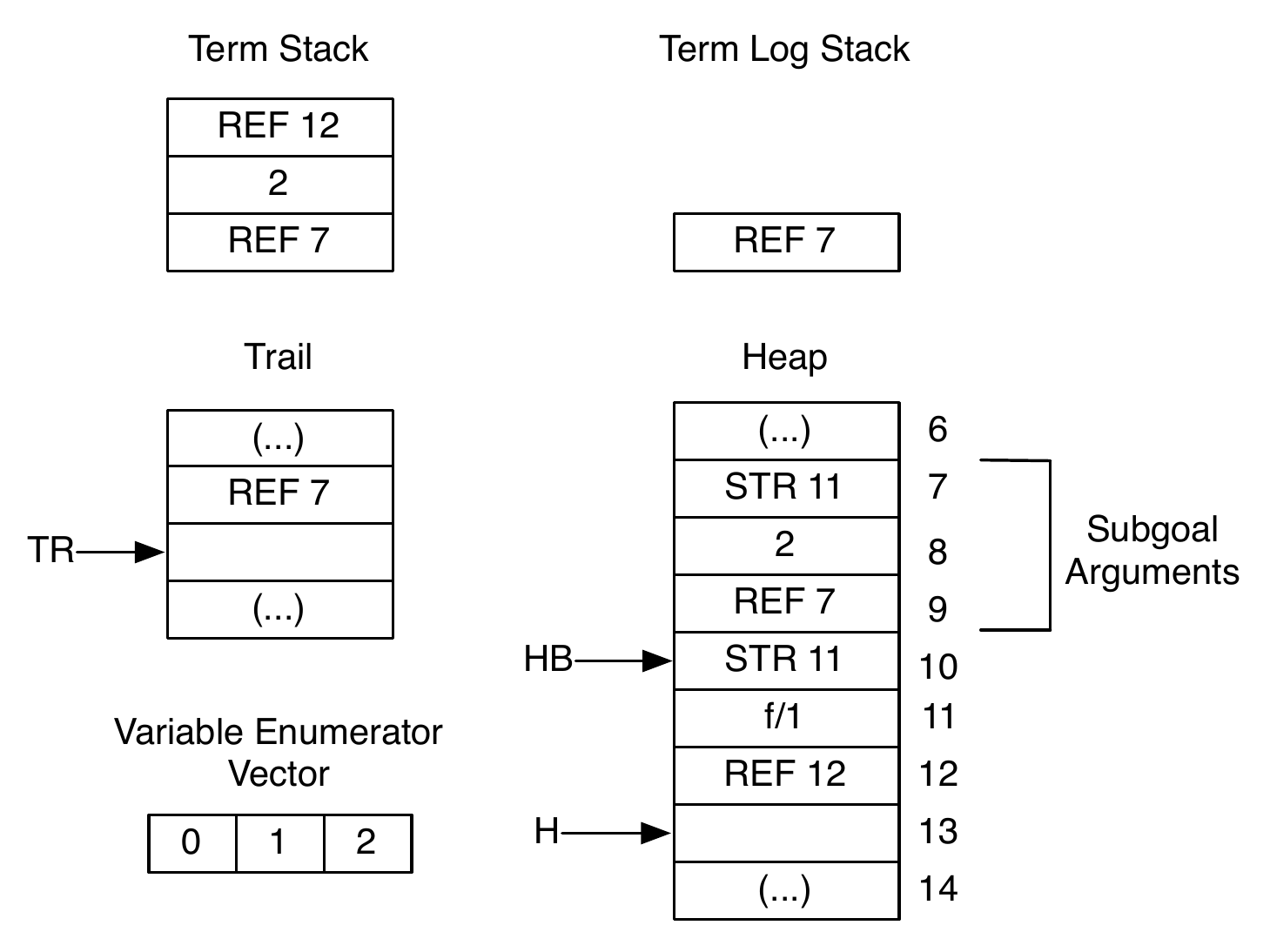}\label{fig:data_structs1}}
\subfigure[After descending into node $n7$]{\includegraphics[scale=0.42]{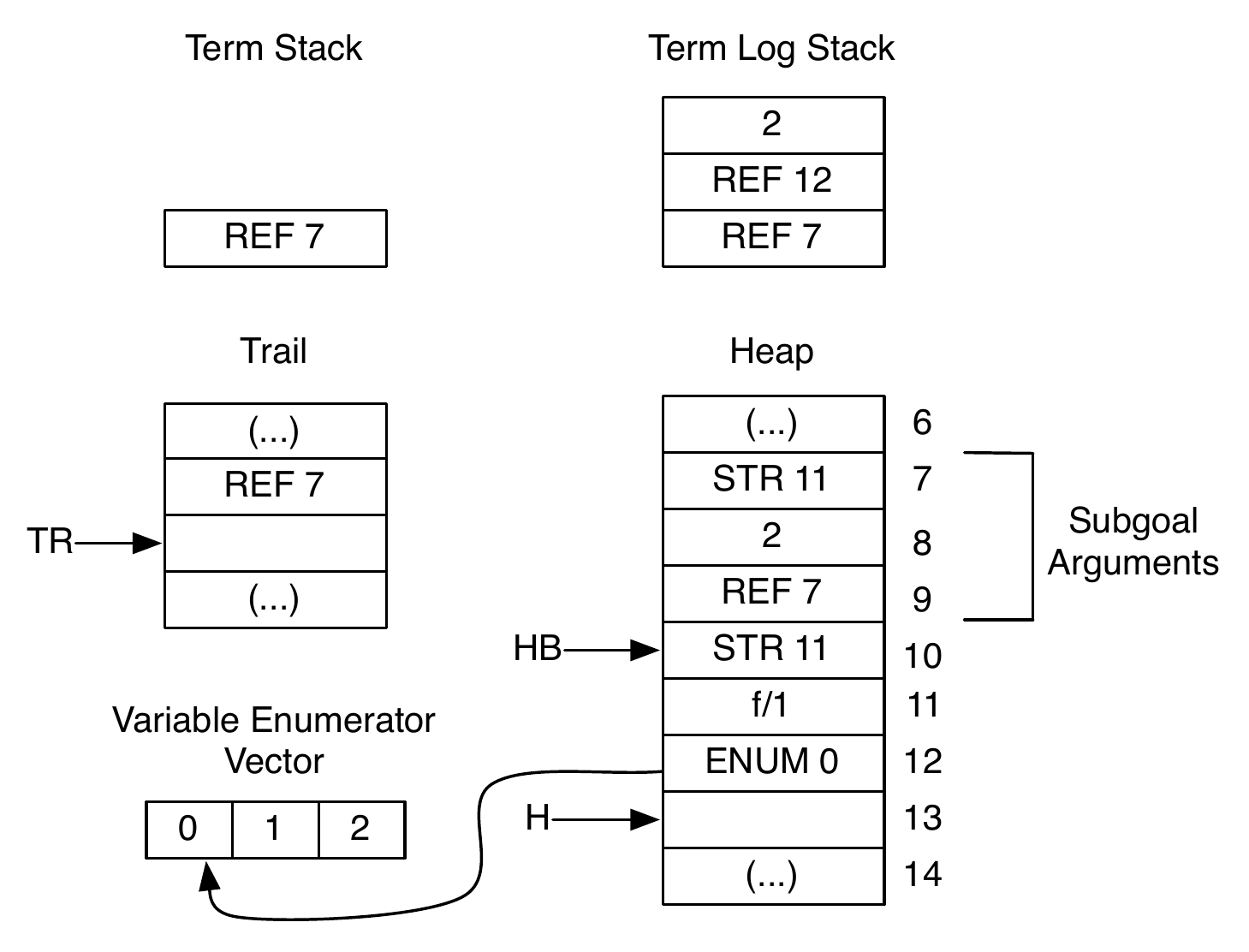}\label{fig:data_structs2}}
\caption{Auxiliary data structures configuration}
\label{fig:data_structs}
\end{figure}

Initially, the algorithm sets up the WAM registers and then pushes the
subgoal arguments into the term stack, resulting in the following
stack configuration (from bottom to top): $[X,2,X]$. Next, we pop the
$X$ variable from the term stack and inspect the linked list of nodes
$n1$, $n9$ and $n12$. Because $X$ is a variable term, we can
potentially match this term with any node with $in\_eval > 0$. We thus
match $X$ against the trie symbol $f/1$ from node $n1$ by constructing
a new $f/1$ functor term into the heap and by binding $X$ to it (this
includes trailing $X$'s reference). Figure~\ref{fig:data_structs1}
shows the configuration of the auxiliary data structures at this
point. Notice that the $H$ register now points to the next free heap
cell, the $X$ variable ($REF~7$) was pushed into the term log stack,
and the new variable representing the argument of $f/1$ ($REF~12$) was
pushed into the term stack. Before descending into node $n2$, we need
to push the alternative node $n12$ into the choice point stack. Note
that node $n9$ cannot be used as an alternative because no evaluating
subgoals exist in that trie branch.

Next, we pop the unbound functor argument from the term stack and we
match it against the trie symbol 3 from node $n2$. Node $n5$ is then
pushed into the choice point stack, and we now have the following
choice point stack configuration: $[n12,n5]$. We then descend into node $n3$, where
the next term from the term stack, 2, matches the trie symbol 2. Here,
there are no alternative nodes to explore, and matching proceeds to
node $n4$. In node $n4$, we first pop the $X$ variable from the term
stack that, after dereferenced, points to the constructed $f/1$
functor on the heap. As we cannot match ground terms with trie
variables, the process fails.

We then pop the top frame from the choice point stack and search is
resumed at node $n5$. The resulting choice point stack configuration
is $[n12]$. Because we backtracked to try node $n5$, the term stack is
restored with the $X$ variable ($REF~7$), the constant 2 and the
functor argument ($REF~12$). In node $n5$, the functor argument is
popped from the term stack and, because the trie symbol is the $VAR_0$
trie variable, it is made to point to the index 0 of the variable
enumerator vector. We then descend into node $n6$, where matching
succeeds and then we arrive at node
$n7$. Figure~\ref{fig:data_structs2} shows the configuration of the
auxiliary data structures at this point. Notice that the twelfth cell
on the heap now points to a variable enumerator position and that the
$HB$ register points again to the tenth cell on the heap, which
corresponds to the value of the $H$ register when we pushed the choice
point frame for the alternative node $n12$.

Node $n7$ contains the functor $f/1$ and the next term from the term
stack is the $X$ variable ($REF~7$) that is bound to the functor $f/1$
created on the heap. Matching therefore succeeds and we get into node
$n8$. In node $n8$, we have the trie variable $VAR_0$ and a variable
that is in the variable enumerator vector. Because they both
correspond to the same index, index 0, matching succeeds and a first
subsumed subgoal is found: $p(f(X),2,f(X))$.

Next, we pop the next top frame from the choice point stack and search
is resumed at node $n12$. The term stack is restored to its initial
state and the choice point stack is now empty. Node $n12$ contains the
trie symbol 5 that matches the term variable $X$. Execution proceeds
to node $n13$ where the trie symbol 2 matches the constant term 2. We
then descend again, now to node $n14$. Here, only node $n15$ can be
used. We then pop the variable $X$ from the term stack that is bound
to the constant 5. As it matches the trie symbol 5 in node $n15$, a
new subsumed subgoal is thus found: $p(5,2,5)$. Finally, as there are
no more alternatives, the algorithm ends and returns the two subsumed
subgoals found.


\section{Experimental Results}

Experiments using the YapTab tabling engine with support for RCS
showed very good results when compared to non-retroactive call by
subsumption~\cite{Cruz-10}. Here, we focus on the retrieval algorithm
and for this we analyze experimental results using some well-known
tabled programs. The environment for our experiments was a PC with a
2.0 GHz Intel Core 2 Duo CPU and 2 GBytes of memory running MacOSX
10.6.6 with YapTab 6.03.

For a better understanding of our algorithm, here called \emph{EIRS
  (Efficient Instance Retrieval of Subgoals)}, we compare it to two
alternative implementations. The first alternative implementation is
called \emph{NIRS (Naive Instance Retrieval of Subgoals)}. It uses the
same matching rules as the main algorithm presented here, but, instead
of using backtracking and pruning by active generators through the
$in\_eval$ field, it starts from a list of subgoal trie leaf nodes,
built during runtime, and attempts to find subsumed subgoals by
matching each subgoal separately in a bottom-up fashion. The second
alternative implementation is called \emph{SIRS (Semi-naive Instance
  Retrieval of Subgoals)} and works exactly like the algorithm
presented in this paper, except that the subgoal trie is not extended
with the $in\_eval$ field, so there is no pruning by inactive
generators.

Table~\ref{tbl:eirs_vs_nirs_and_sirs} shows execution statistics for
EIRS, NIRS and SIRS when running some well-known tabled
programs\footnote{All programs available
  from~\url{http://cracs.fc.up.pt/node/4695}}. The first column shows
the program name and the second column shows the number of calls to
the retrieval algorithm. Then, for each algorithm, we present the
execution time, in milliseconds, that was spent executing the
retrieval of subsumed subgoals (average of three runs) and, between
parenthesis, we include the percentage of time that this task
represents in the execution of the whole program. Note that for EIRS,
the execution time also includes the maintenance of the $in\_eval$
fields and related data structures. Finally, we
show the speedup of EIRS over NIRS and SIRS (values in bold mean that EIRS is better).

\begin{table}[ht]
\caption{Execution statistics for EIRS, NIRS and SIRS}
\begin{tabular}{lcrrrrr}
\hline\hline
\multirow{2}{*}{\textbf{Program}} & 
\multicolumn{1}{c}{\multirow{2}{*}{\textbf{Calls}}} & 
\multicolumn{1}{c}{\multirow{2}{*}{\textbf{EIRS}}} &
\multicolumn{1}{c}{\multirow{2}{*}{\textbf{NIRS}}} &
\multicolumn{1}{c}{\textbf{NIRS}} &
\multicolumn{1}{c}{\multirow{2}{*}{\textbf{SIRS}}} &
\multicolumn{1}{c}{\textbf{SIRS}} \\ \cline{5-5} \cline{7-7}
& & & & \multicolumn{1}{c}{\textbf{EIRS}} & & \multicolumn{1}{c}{\textbf{EIRS}} \\
\hline
{\bf empty (100)} &   101 &   0.54 (81)  &     0.22 (22) &        0.41  &  0.71 (71)  &   {\bf 1.32} \\
{\bf empty (1K)}  &  1001 &   5.33 (10)  &     7.32 (12) &   {\bf 1.37} &  7.03 (12)  &   {\bf 1.32} \\
{\bf empty (10K)} & 10001 &  54.50 (0.9) &   668.66 (11) &  {\bf 12.30} & 71.15 (1.2) &   {\bf 1.31} \\
{\bf one (100)}   &   102 &   0.55 (82)  &     0.23 (35) &        0.42  &  0.71 (71)  &   {\bf 1.29} \\ 
{\bf one (1K)}    &  1002 &   5.42 (7)   &     7.55 (9)  &   {\bf 1.40} &  7.32 (9.5) &   {\bf 1.35} \\
{\bf one (10K)}   & 10002 &  54.50 (0.4) &   710.50 (5)  &  {\bf 13.04} & 73.97 (0.5) &   {\bf 1.36} \\
{\bf end (100)}   &   101 &   0.58 (87)  &     1.75 (88) &   {\bf 3.01} &  0.71 (71)  &   {\bf 1.22} \\
{\bf end (1K)}    &  1001 &   5.92 (10)  &   143.77 (70) &  {\bf 24.29} &  7.02 (11)  &   {\bf 1.19} \\
{\bf end (10K)}   & 10001 &  59.69 (0.9) & 17361.50 (59) & {\bf 290.90} & 71.21 (1)   &   {\bf 1.19} \\
\hline                                                                                                      
{\bf flora}       &   226 &   1.39 (1)   &    59.55 (24) &  {\bf 42.93} & 24.41 (15)  &  {\bf 17.34} \\
{\bf genome1}     &     4 &   0.02 (0)   &    13.95 (23) & {\bf 606.52} & 14.14 (23)  & {\bf 614.78} \\
{\bf genome2}     &     4 &   0.02 (0)   &     6.86 (25) & {\bf 285.83} &  7.08 (24)  & {\bf 354.10} \\
{\bf genome3}     &     4 &   0.02 (0)   &     3.52 (12) & {\bf 146.66} &  3.53 (12)  & {\bf 147.08} \\
\hline\hline
\end{tabular}
\label{tbl:eirs_vs_nirs_and_sirs}
\end{table}

The \textbf{empty} programs consist in the evaluation and completion
of several subgoals (total number in parentheses) followed, in the
end, by a call to a more general subgoal. The results for the
\textbf{empty} programs show that, as the number of subgoals
increases, both EIRS and SIRS perform in linear time, while the NIRS
does not keep up and behaves in a super-linear fashion. However, EIRS
tends to be 30\% faster than SIRS. The \textbf{one} programs perform a
search for a single subsumed subgoal in a subgoal trie where several
subgoals have already completed (total number in parentheses). The
\textbf{one} programs have a comparable behavior to that of
\textbf{empty}. The \textbf{end} programs consist in the evaluation of
several subgoals (total number in parentheses) followed by a call to a
subgoal that subsumes every other in evaluation. The results for the
\textbf{end} programs show that both EIRS and SIRS keep their $O(n)$
complexity for this benchmark, while NIRS performs very badly. Here,
EIRS appears to be 20\% better than SIRS, which can be explained by
the need of a new stack to control the traversal of the hash table,
for the case when the next term to match is a variable.

The second section of Table~\ref{tbl:eirs_vs_nirs_and_sirs} presents
real-world examples of tabled programs. We included a relatively
complex benchmark, \textbf{flora}, from the Flora object-oriented
knowledge base language~\cite{Yang-00}, which shows a 42-fold and
17-fold speedup over NIRS and SIRS, respectively. If we take into
account that \textbf{flora} takes 165 milliseconds to run the entire
program with EIRS, the fraction of time spent retrieving subsumed
subgoals is very small (less than 1\%). However, if we use NIRS or
SIRS, the fraction of time spent collecting subgoals is 24\% or 15\%,
which is very considerable.  The \textbf{genome} programs also show
the same behavior and even more impressive speedups for the EIRS
algorithm.

In summary, the results in Table~\ref{tbl:eirs_vs_nirs_and_sirs}
clearly show that the EIRS backtracking mechanism for trie traversal
is quite efficient and effective as the number of subgoals
increases. The NIRS algorithm performs worse in most cases, since it
needs to match every subgoal separately. The SIRS algorithm tends to
perform better than NIRS in our synthetic programs, but, in some
important benchmarks, like \textbf{flora} or \textbf{genome}, the
results are not so good, since the algorithm is unable to ignore trie
branches where no active generators exist.


\section{Conclusions}

We presented a new algorithm for the efficient retrieval of subsumed
subgoals in tabled logic programs. Our proposal takes advantage of the
existent WAM machinery and data areas and extends the subgoal trie
data structure with information about the evaluation status of the
subgoals in a branch, which allows us to reduce the search space
considerably. We therefore argue that our approach can be easily
ported to other tabling engines, as long they are based on WAM
technology and use tries for the table space.

Our experiments with the retrieval algorithm presented in this paper
show that the algorithm has good running times, even when the problem
size is increased, thus evidencing that our approach is robust and
capable of handling programs where intensive retroactive subsumption
is present.


\section*{Acknowledgments}

This work has been partially supported by the FCT research projects
HORUS (PTDC/EIA-EIA/100897/2008) and LEAP (PTDC/EIA-CCO/112158/2009).


\bibliographystyle{acmtrans}
\bibliography{references}

\begin{thebibliography}{}

\bibitem[\protect\citeauthoryear{Chen and Warren}{Chen and
  Warren}{1996}]{Chen-96}
{\sc Chen, W.} {\sc and} {\sc Warren, D.~S.} 1996.
\newblock {Tabled Evaluation with Delaying for General Logic Programs}.
\newblock {\em Journal of the ACM\/}~{\em 43,\/}~1, 20--74.

\bibitem[\protect\citeauthoryear{Cruz and Rocha}{Cruz and
  Rocha}{2010}]{Cruz-10}
{\sc Cruz, F.} {\sc and} {\sc Rocha, R.} 2010.
\newblock {Retroactive Subsumption-Based Tabled Evaluation of Logic Programs}.
\newblock In {\em European Conference on Logics in Artificial Intelligence}.
  Number 6341 in LNAI. Springer-Verlag, 130--142.

\bibitem[\protect\citeauthoryear{Johnson}{Johnson}{2000}]{Johnson-00}
{\sc Johnson, E.} 2000.
\newblock {Interfacing a Tabled-WAM Engine to a Tabling Subsystem Supporting
  Both Variant and Subsumption Checks}.
\newblock In {\em Conference on Tabulation in Parsing and Deduction}. 155--162.

\bibitem[\protect\citeauthoryear{Johnson, Ramakrishnan, Ramakrishnan, and
  Rao}{Johnson et~al\mbox{.}}{1999}]{Johnson-99}
{\sc Johnson, E.}, {\sc Ramakrishnan, C.~R.}, {\sc Ramakrishnan, I.~V.}, {\sc
  and} {\sc Rao, P.} 1999.
\newblock {A Space Efficient Engine for Subsumption-Based Tabled Evaluation of
  Logic Programs}.
\newblock In {\em Fuji International Symposium on Functional and Logic
  Programming}. Number 1722 in LNCS. Springer-Verlag, 284--300.

\bibitem[\protect\citeauthoryear{Ramakrishnan, Rao, Sagonas, Swift, and
  Warren}{Ramakrishnan et~al\mbox{.}}{1999}]{RamakrishnanIV-99}
{\sc Ramakrishnan, I.~V.}, {\sc Rao, P.}, {\sc Sagonas, K.}, {\sc Swift, T.},
  {\sc and} {\sc Warren, D.~S.} 1999.
\newblock {Efficient Access Mechanisms for Tabled Logic Programs}.
\newblock {\em Journal of Logic Programming\/}~{\em 38,\/}~1, 31--54.

\bibitem[\protect\citeauthoryear{Rao, Ramakrishnan, and Ramakrishnan}{Rao
  et~al\mbox{.}}{1996}]{Rao-96}
{\sc Rao, P.}, {\sc Ramakrishnan, C.~R.}, {\sc and} {\sc Ramakrishnan, I.~V.}
  1996.
\newblock {A Thread in Time Saves Tabling Time}.
\newblock In {\em Joint International Conference and Symposium on Logic
  Programming}. The MIT Press, 112--126.

\bibitem[\protect\citeauthoryear{Rao, Sagonas, Swift, Warren, and Freire}{Rao
  et~al\mbox{.}}{1997}]{Rao-97}
{\sc Rao, P.}, {\sc Sagonas, K.}, {\sc Swift, T.}, {\sc Warren, D.~S.}, {\sc
  and} {\sc Freire, J.} 1997.
\newblock {XSB: A System for Efficiently Computing Well-Founded Semantics}.
\newblock In {\em International Conference on Logic Programming and
  Non-Monotonic Reasoning}. Number 1265 in LNCS. Springer-Verlag, 431--441.

\bibitem[\protect\citeauthoryear{Rocha, Silva, and {Santos Costa}}{Rocha
  et~al\mbox{.}}{2000}]{Rocha-00a}
{\sc Rocha, R.}, {\sc Silva, F.}, {\sc and} {\sc {Santos Costa}, V.} 2000.
\newblock {YapTab: A Tabling Engine Designed to Support Parallelism}.
\newblock In {\em Conference on Tabulation in Parsing and Deduction}. 77--87.

\bibitem[\protect\citeauthoryear{Rocha, Silva, and {Santos Costa}}{Rocha
  et~al\mbox{.}}{2005}]{Rocha-05a}
{\sc Rocha, R.}, {\sc Silva, F.}, {\sc and} {\sc {Santos Costa}, V.} 2005.
\newblock {On applying or-parallelism and tabling to logic programs}.
\newblock {\em Theory and Practice of Logic Programming\/}~{\em 5,\/}~1 \& 2,
  161--205.

\bibitem[\protect\citeauthoryear{Yang and Kifer}{Yang and
  Kifer}{2000}]{Yang-00}
{\sc Yang, G.} {\sc and} {\sc Kifer, M.} 2000.
\newblock {Flora: Implementing an Efficient Dood System using a Tabling Logic
  Engine}.
\newblock In {\em Computational Logic}. Number 1861 in LNCS. Springer-Verlag,
  1078--1093.

\end{thebibliography}


\end{document}